\documentclass[preprint2]{emulateapj}

\usepackage{amssymb}
\usepackage{multirow}

\usepackage{graphicx}

\def\cm{\,{\rm cm}}

\def\ergscm2 {erg\,s$^{-1}$cm$^{-2}$}

\def\cm2 {cm$^{-2}$}

\def\aap {A\&A}
\def\apj {ApJ}

\shorttitle{QNe in massive binaries : A universal energy source for double-peaked SLSNe}
\shortauthors{Ouyed et al.}

\begin{document}
 
\title{ Quark-Novae occurring in massive binaries : A universal energy source in superluminous Supernovae with double-peaked light curves}
 
\author{Rachid Ouyed\thanks{Email:rouyed@ucalgary.ca}, Denis Leahy and Nico Koning}

\affil{Department of Physics \& Astronomy, University of Calgary, 2500 University Drive NW, Calgary, AB T2N 1N4, Canada}

\begin{abstract} 
 A Quark-Nova (QN;  the sudden transition from a neutron star into a quark star) which occurs in the second common envelope (CE) phase of a massive binary (Ouyed et al., 2015a\&b), gives excellent fits to super-luminous, hydrogen-poor, Supernovae (SLSNe)  with double-peaked light curves including DES13S2cmm, SN 2006oz and LSQ14bdq ({\it http://www.quarknova.ca/LCGallery.html}). In our model,  the H envelope of the less massive companion is ejected during the first CE phase  while the QN occurs deep inside the second, He-rich, CE phase  after the CE has expanded in size to a radius of a few tens to a few thousands solar radii; this yields the  first peak in our model. The ensuing merging of the quark star with the CO core leads to black hole formation and accretion explaining the second long-lasting peak. 
 We study a sample of 8 SLSNe Ic with double-humped light-curves. Our model provides good fits to all of these with
a universal explosive energy of $2\times 10^{52}$ erg (which is the kinetic energy of the QN ejecta) for the first hump. The late-time emissions seen in iPTF13ehe and LSQ14bdq are fit with a shock interaction between the outgoing He-rich (i.e second) CE and the previously ejected H-rich (i.e first) CE.
 \end{abstract}

\keywords{Circum-stellar matter -- stars: evolution -- stars: winds, outflows -- supernovae: general -- supernovae: individual (iPTF13ehe, LSQ14bdq)}


\section{Introduction}

 Super-luminous, H-poor, supernovae (SLSNe-I) occur with different-shaped light-curves and lack H in their spectra (\cite{postorello_2010,quimby_2011,inserra_2013}). Their peak luminosities are more than a factor 10 higher than for Type-Ia SNe and for other types of core-collapse SNe (\cite{galyam_2012}).  SLSNe-I occur in low-metallicity environments/galaxies with properties that are unlike what is expected in other core-collapse SNe and or in Type-Ia SNe
(\cite{chen_2013}; \cite{nicholl_2015a}; \cite{leloudas_2015}; \cite{lunann_2015}).
 The mechanisms and the progenitors of these SLSNe-I are still unclear and debated,
 with powering by black hole (BH) accretion (\cite{dexter_2013}) and  millisecond magnetars (\cite{woosley_2010, kasen_2010}) suggested.  Other models appeal to the interaction of  SN ejecta with its surroundings (\cite{wang_2015,moriya_2015,sokorina_2015,piro_2015}) as well as a binary origin (\cite{moriya_2015}).

In \cite{ouyed_2015a,ouyed_2015b} we showed  how SLSNe-I light-curves, such as for LSQ14bdq (\cite{nicholl_2015b}), SN 2006oz (\cite{leloudas_2012})
and DES13S2cmm (\cite{papadopoulos_2015}) can be  fit by a QN in a massive stellar binary. Our binary model (illustrated in Figure 1 in \cite{ouyed_2015a}), has component A (i.e. the more massive component) explode as a normal SN, which produces a Neutron Star (NS), which then ejects the H envelope of its companion star when the binary begins a first CE phase. In the second CE phase, the NS spirals-in closer and
expands a second He CE to a  radius of a few tens to a few thousands solar radii. The accreting NS
 gains sufficient mass to undergo a QN explosion leaving behind a quark star (QS).
The energization of the He-rich (i.e. second envelope) by the QN yields  the first light-curve peak. The QN inputs sufficient energy  ($\sim 2\times10^{52}$ erg of kinetic energy in the QN ejecta) to eject the He-rich (i.e. second) envelope. Further orbital decay with the following merging of the QS and the CO core of the companions star B, changes the QS into a BH. BH-accretion provides the power for the second bright/long-lasting peak,
which is set by the mass of the CO core. The eventual interaction of the QN-BH-energized He-rich
CE with the previously ejected H-rich envelope (i.e. the first CE) can yield a late-time brightening in the light-curve,
with emergence of Hydrogen in the spectrum. This has been reported
in the late-time spectra of the recently discovered SLSN-I iPTF13ehe (\cite{yan_2015}). This is suggestive of QNe occurring in massive binaries as described in our model.

Here we demonstrate that a QN in a massive binary which undergoes two CE phases \cite{ouyed_2015a,ouyed_2015b} fits the light-curves of  many hydrogen-poor double-humped SLSNe. We find that the first peak in these SLSNe can be fit with a constant energy injection (i.e. the $2\times 10^{52}$ of QN ejecta kinetic energy) into different sizes of the He-rich, second CE.  The paper is structured as follows: in \S 2, an overview of the QN model and QNe in binaries is given. In \S 3, the results of applying the model to 8 SLSNe-I is given, including
iPTF13ehe and LSQ14bdq  which show clear late-time emission. We conclude in \S 4.

\section{The Quark-Nova in a massive binary}

  \subsection{The QN Energetics}

  With the assumption that matter made of up, down and strange quarks (hereafter {\it uds} matter) is more stable than
  hadronic matter (\cite{bodmer_1971,witten_1984}),  the conversion of a NS to a {\it uds}  quark star is a tantalizing possibility. The QN is the {\it explosive}  conversion of a NS to a quark star (QS) (\cite{ouyed_2002}; see  \cite{ouyed_2013} for a recent review).  If the NS is massive enough, quark deconfinement can occur in its dense core which when combined with t  strange-quark seeding can trigger the conversion.   We define the critical value
of the NS mass, $M_{\rm NS, c.}$, which is taken as $2M_{\odot}$ to account for massive NSÕs (\cite{demorest_2010}).

  Analytical and numerical  investigations of this  transition
  can be found in the literature   with differing outcomes (to cite only a few \cite{horvath_1988,ouyed_2002,keranen_2005,niebergal_2010, herzog_2013,pagliara_2013,furusawa_2015a,furusawa_2015b,drago_2015a}).
 More  recent investigations suggest   three paths to  ejection of the outer layers of the NS. First are the
     neutrino-driven explosion mechanism where the energy is deposited  at the bottom of the NS crust (\cite{keranen_2005,drago_2015b}). The   neutrinos  emitted from the conversion to {\it uds} 
transport the energy into the outer regions of the NS star, leading
to heating and subsequent mass ejection. This is unlike a Supernova explosion where most of the neutrinos escape
      from the system.  For a density of a few times nuclear saturation density inside the NS and a neutrino capture cross-section of $\sim 10^{-43}$-$10^{-42}$ cm$^2$,  the neutrino mean-free-path is of the order of a few hundreds of centimeters. This means that  at most 10\% are lost  to the system. Second, in addition to shock and neutrino heating,  the transition yields a copious amount  of photons as the {\it uds} cools and enters a superconducting
     state (\cite{alford_1998}). In this regime  the quarks pair and  photon emissivity becomes important (\cite{vogt_2004,ouyed_2005}) generating a photon fireball which induces a thermal ejection/explosion of the NS outer layers;  A third avenue consists of a quark-core collapse explosion possibly induced by the
      de-leptonization instability (\cite{niebergal_2010,ouyed_2013}).

         The general picture is that the (neutrino and photon) fireball from the QN coupled
     with the core-collapse acts as a piston at the base of the crust, i.e., the QS fireball expands
approximately adiabatically while pushing the overlaying crust, and cooling fairly rapidly.  The fireball imparts to 
     the NS outer layers ($\sim 10^{-3}M_{\odot}$) about $2\times 10^{52}$ ergs  in kinetic energy.   Other models  appeal to a two-step conversion of nuclear matter to {\it uds} quark matter yielding similar energies (\cite{dai_1995})
   with applications to gamma-ray bursts (\cite{cheng_1996}).  From more general principles one can estimate the energy release   by noting that the conversion of a hadron to its constituent quarks yields about $E_{\rm conv.}\sim10^{-4}$ ergs  in energy (e.g. \cite{weber_2005}).  A QN can thus  release about $(M_{\rm NS, c.}/m_{\rm H}) E_{\rm conv.}\sim 2\times 10^{53}\ {\rm ergs}\ (M_{\rm NS, c.}/2M_{\odot})$  from the direct conversion alone without accounting for the additional energy released as photons and as gravitational binding energy. The extremely
dense environment involved means that at least $\sim 10$\% of this energy is converted to kinetic energy of the QN ejecta. On average, neutron-rich material of mass $M_{\rm QN} \sim 10^{-3}$ M$_\odot$,  is ejected during the QN (\cite{keranen_2005, ouyed_2009, niebergal_2010}) with a Lorentz factor $\Gamma_{\rm QN}\sim 10$, which corresponds to a kinetic energy of $E_{\rm QN}\sim 2\times 10^{52}$ erg.

   \subsection{The binary}
  
  In our scenario, the QN event occurs in a binary where  the NS (formed from the primary) accretes from
secondary to reach the critical mass ($M_{\rm NS, c.}$) above which the QS conversion is triggered.
The events that lead to the QN have been described in \S 3 (and related Figure 1) in \cite{ouyed_2015a}, which starts with a high mass binary. The higher-mass component (A) undergoes a SN explosion creating a NS.
The subsequent evolution of the binary is dictated by the evolution of component B which expands and
engulfs the NS when its CO core has developed. This triggers a CE phase which results in  ejection of the hydrogen (H) envelope of component B by the NS. Thus following this first CE phase, the system is left as a NS binary in a close orbit
(i.e. a period of hours) with the stripped He-rich component B.

  Before we describe the subsequent stages of evolution towards the second CE phase and its ejection,
  we note that the timescale for the growth of a CO core in component B is $\tau_{\rm B, CO}\sim M_{\rm CO} \epsilon_{\rm CO}/L_{\rm He}\sim 10^5\ {\rm years}$
where $M_{\rm CO}$ is the mass of the CO core of  B, $\epsilon_{\rm CO}\sim 5.9\times 10^{17}$ erg g$^{-1}$  the energy release per unit mass and $L_{\rm He}\sim  10^5L_{\odot}$
is the He burning luminosity. The star's radius when it has a significant CO core ($\sim 2M_{\odot}$) is  $\sim 1000R_{\odot}$ (e.g. \cite{salaris_2005}). This means for the first CE phase to start when the CO core of B  is already formed we require an initial orbital separation\footnote{The percentage of massive binary stars with orbital separation exceeding $\sim 1000R_{\odot}$ is of the order of 10-15\% (\cite{kobulnicky_2014}). For systems with smaller initial orbital separation 
(between the NS and the B component), the lack of well defined CO core   (i.e. sharp density gradient at the edge
of the core)  means that the second CE will fully eject the companion leaving behind either a QS or a BH
without a substantial CO core to accrete from. In this case we do not expect a BH accretion phase (i.e. the second hump). However we do expect the CSM interaction.} exceeding $\sim 1000R_{\odot}$.

  \subsection{The QN phase}
   \label{sec:qn}
  
  After ejection of the H-rich CE,
the secondary star B evolves while expanding. This causes the second CE phase which triggers additional accretion onto the NS. During this period, the envelope expands
to large radii\footnote{The frictional energy released during orbital shrinkage is transmitted by convection and then radiated without envelope ejection (\cite{meyer_1979}; see \S 2.2 in \cite{ouyed_2015a} on how this steady energy balanced is reached ).}. The time it would take the envelope to inflate to  a radius of hundreds of solar radii is of the order of a few years, which is  close to the time needed for the NS to accrete enough mass to reach $M_{\rm NS, c.}$ and undergo a QN explosion. This resulting release of $\sim 2\times 10^{52}$ erg of QN kinetic energy, is  deposited into the second CE, and yields the bright, first peak, in the light-curve. The energy release from the QN is sufficient to unbind the He-rich CE then eject it at a speed $v_{\rm QN}$ of order a few $10,000$ km s$^{-1}$. As we argued in our previous
papers, in this scenario (involvinh a He-rich second CE), the QN induces CE ejection. After the ejection of the He-rich (i.e. the second) CE, the system is left as a $\sim 2M_{\odot}$ QS in orbit with a CO core
of mass $M_{\rm CO} \sim 2M_{\odot}$ and radius $R_{\rm CO} < 0.1R_{\rm sun}$.

The computation of the QN light curve is given in  Appendix in \cite{ouyed_2015b} where the He-rich CE
initial shock temperature ($T_{\rm CE, 0}$; following impact by the QN ejecta) was estimated from the shocked gas pressure. Here we provide a more complete  prescription using radiative shock jump conditions as follows:
 (i) The total  energy of the shocked  CE is $E_{\rm QN}=  E_{\rm CE, KE}+ E_{\rm CE, th.}$ with 
 $E_{\rm CE, KE}= \frac{1}{2} M_{\rm CE}v_{\rm QN}^2$ being the kinetic energy of the envelope and $E_{\rm CE, th.} = N_{\rm CE} c_{\rm V} T_{\rm CE, 0}+ a T_{\rm CE, 0}^4 V_{\rm CE, 0}$ its thermal energy (gas and radiation). Here, $c_{\rm V}$ is the specific heat, 
 $\mu_{\rm He}$ the CE's mean molecular weight,  $a$ the radiation constant, 
  $N_{\rm CE}= M_{\rm CE}/(\mu_{\rm He} m_{\rm H})$ and $V_{\rm CE, 0}= (4\pi/3) R_{\rm CE, 0}^3$
   the CE initial volume;   (ii) For high-velocity shocks occurring in low-density gas as in our case, the downstream conditions are dominated by radiation pressure. The shock jump conditions yield $a T_{\rm shock}^4 =  (6/7) \rho_{\rm CE} v_{\rm shock}^2$ for a CE density $\rho_{\rm CE}$ with  a density compression ratio of 7; $v_{\rm shock}$ and $T_{\rm shock}$ are the shock velocity and temperature, respectively.  This implies that  the downstream velocity (in our case $v_{\rm QN}$) is $v_{\rm QN}= (6/7) v_{\rm shock}$.  The high compression ratio also means the post-shock material (in the He-rich CE) is in a thin shell with volume 1/7 of its original volume; (iii) The sound speed for the hot thin shell  is larger than $v_{\rm QN}$ so that the shell (i.e. the shocked part of the He-rich CE)
can quickly expand to refill its original spherical volume.  Because the shell is optically thick  it expands
adiabatically and remains radiation dominated. The resulting temperature of the hot interior is $7^{-1/3} T_{\rm shock}$.
 This interior temperature is what we take to be the initial temperature of the spherical QN-shocked CE, $T_{\rm CE, 0} = 7^{-1/3}\times (7 \rho_{\rm CE}  v_{\rm QN}^2/6a)^{1/4}$; 
 (iv) Thus, given  $E_{\rm QN} = 2\times 10^{52}$ ergs (fixed) both $T_{\rm CE, 0}$ and $v_{\rm QN}$ (constant in our model) are determined and the  only parameters we vary are  the mass of the He-rich (i.e. second) CE ($M_{\rm He}$)
 and its initial radius $R_{\rm CE, 0}$.  The evolution of the CE radius is then
 $R_{\rm CE}= R_{\rm CE, 0} + v_{\rm QN} (t-t_{\rm BO})$ where $t_{\rm BO}$ is the time
 of QN shock breakout; see Appendix in \cite{ouyed_2015b} on the time evolution of
 the CE radius, its temperature, the photosphere  and the calculation of the QN light curve.

  \subsection{The BH accretion phase}
   \label{sec:bh}

The orbital separation at the time QN event is of order $\sim 0.5 R_{\odot}$ (see \S 3 in \cite{ouyed_2015a}).
The statement in \cite{ouyed_2015a} that gravitational radiation can explain the time interval between the
QN and the beginning of the BH-accretion phase was incorrect due to a numerical error.  The fits to more
 SLSNe as performed here (see \S \ref{sec:otherslsne}) confirm that a time delay of days to weeks is needed.  Instead we suggest that the Roche-Lobe overflow timescale is a more reasonable explanation for the time delay.  
  This remains to be confirmed by more detailed numerical simulations.

 As the QS accretes from the CO core of component B, it reaches the critical mass to become a BH.  The ensuing BH  accretion re-energizes the expanding He-rich CE and  powers the long lasting main (i.e. second) hump of the light curve. 
The duration of the  BH-accretion phase 
is related to the time it takes the BH to consume the CO core, $\int_{t_0}^{t_0+ \Delta t_{\rm BH}} L_0 (t/t_0)^{-n_{\rm BH}} dt= \eta_{\rm BH} M_{\rm CO} c^2$ which yields, when $n_{\rm BH}\neq 1$,
\begin{equation}
\label{eq:dtBH}
 \Delta t_{\rm BH}  = t_0 \left(     \left((1-n_{\rm BH})\frac{\eta_{\rm BH}M_{\rm CO}c^2}{L_0t_0}+1\right)^{\frac{1}{1-n_{\rm BH}}}   -1  \right)\ ,
\end{equation}
where $\eta_{\rm BH}$ is the accretion efficiency, $L_0$ the BH initial accretion luminosity and $n_{\rm BH}$ a parameter  defining the injection power  $L_0 (t/t_0)^{-n_{\rm BH}}$ (see \cite{dexter_2013}). BH-accretion
turns on at time $t_0$ with $t=0$ corresponding to occurrence of the QN.
The BH accretion phase begins after QN by a time $t_{\rm BH, delay}$, so that $t_0= (R_{\rm CE, 0}/v_{\rm QN})+ t_{\rm BH, delay}$ here.

\begin{figure*}[t!]
\centering
\includegraphics[scale=0.68]{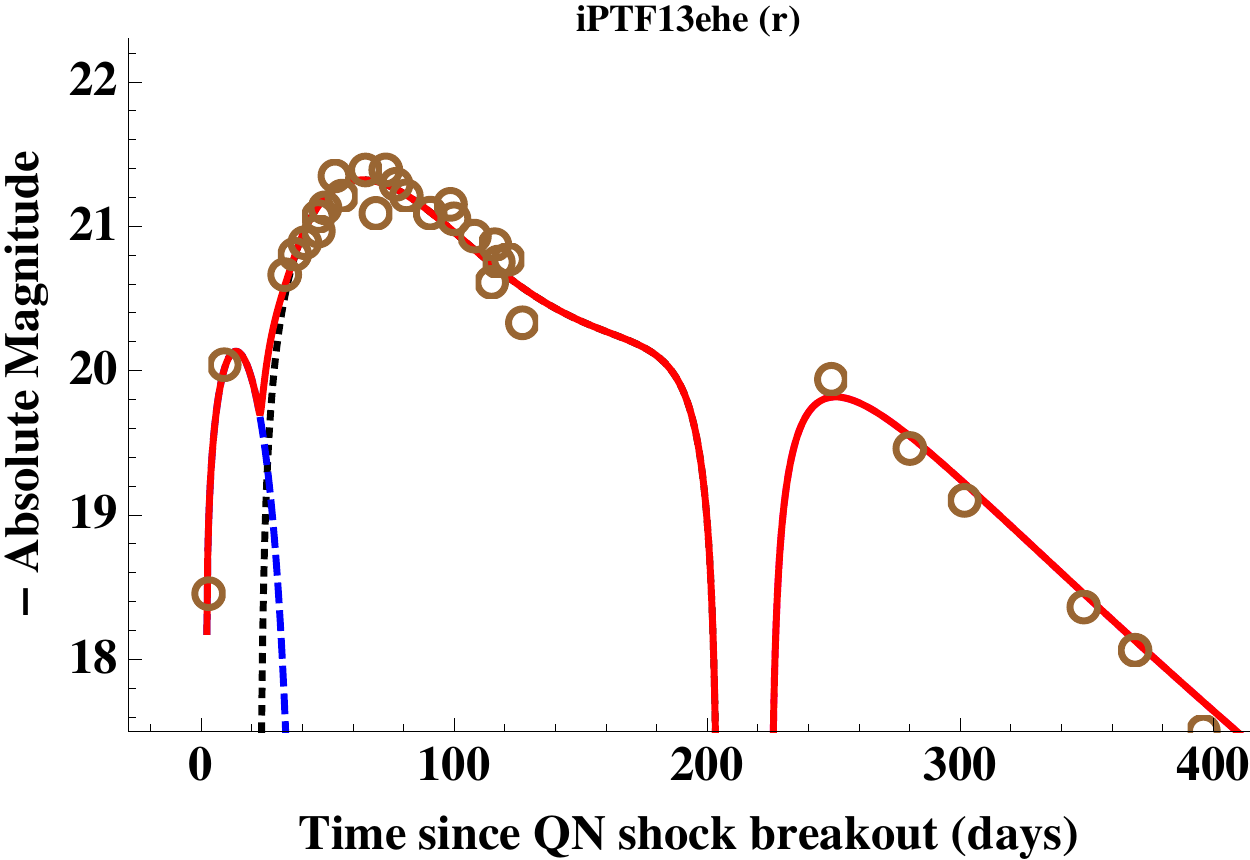}
\includegraphics[scale=0.68]{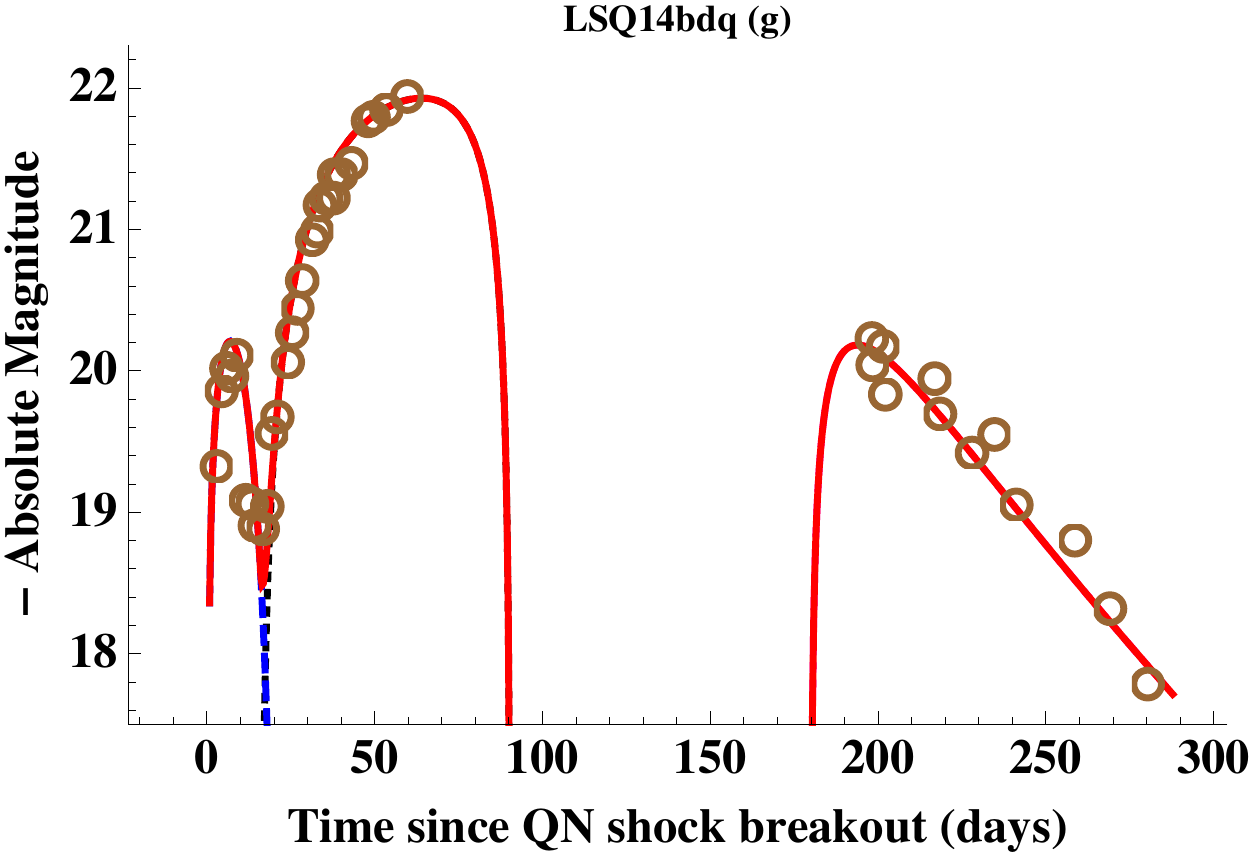}
  \caption{Left Panel: QN model fit (solid line) to the r-band light-curve for iPTF13ehe (Observations (circles) from Yan et al. 2015). Right Panel: QN model fit (solid line) to the g-band light curve of LSQ14bdq (observations (circles) from Nicholl et al. 2015b). For all panels, the blue dashed curve is the QN, the black dotted curve is the BH accretion and the green dot-dashed curve the CSM interaction.}
\label{fig:fits}
\end{figure*}

 \begin{table*}[t!]
\begin{center}
\caption{Best fit parameters  for the  iPTF13ehe and LSQ14bdq  LCs  in our model.}
\begin{tabular}{|c||c|c||c|c|c||c|c|c|c|}\hline
  & \multicolumn{2}{|c||}{QN} &  \multicolumn{3}{|c||}{BH Accretion}  & \multicolumn{4}{|c|}{CSM Interaction$^{\dagger}$}\\
  \hline
SLSN-I  &  $M_{\rm He}$ ($M_{\odot}$) & $R_{\rm CE, 0} (R_{\odot})$    &    $t_{\rm BH, delay}$ (days) & $L_{\rm 0}$ (erg/s) & $n_{\rm BH}$ & $t_{\rm CSM}$ (days)& $\dot{M}_{\rm CE, H}$ $M_{\odot}$ yr$^{-1}$& $v_{\rm w}=v_{\rm H}$ (km/s) & $s$ \\\hline
 iPTF13ehe &   6.0  & 2000   &  22.0  &  $5.5\times 10^{44}$ & 1.1 & 225 & $10^{-2}$ & 180 &   1.75    \\\hline
 LSQ14bdq &  2.8  & 900   &  15.5  &  $1.3\times 10^{44}$ & -0.4 &  190 & $ 10^{-2}$ & 240 &   1.75     \\\hline
\end{tabular}\\
$^{\dagger}$ The CSM  luminosity depends on $D= \dot{M}_{\rm CE, H}/4\pi v_{\rm H} R_{\rm H}^{2-s}$  so
 there are effectively two fit parameters, $t_{\rm CSM}$ and  $\dot{M}_{\rm CE, H}/v_{\rm w}$. The  other CSM parameters, $n=12, \delta =0.0$ and $\epsilon=0.1$ are kept constant   (see \S \ref{sec:csm}).  
 \end{center}
\label{table:params}
\end{table*}

   \subsection{The circumstellar medium (CSM) interaction phase}
   \label{sec:csm}

The subsequent  stage in our model is when the ejected He-rich CE catches up with, and shocks, the H-rich (i.e. first) CE
leading to a  third hump.    The LCs of iPTF13ehe and LSQ14bdq show clear evidence of emission beyond the two peaks which we attribute to such an interaction. I.e. the collision of the re-energized (QN-ejected) second
CE with the surrounding CSM (the H-rich CE material ejected during the first CE phase).

 As illustrated in Figure 1 in \cite{ouyed_2015a}, following the first CE phase, the system is a He-core-NS binary with a separation of order  a few solar radii ($\sim 3R_{\odot}$). The He core has just lost its hydrogen envelope and it is out of equilibrium. It will expand from an initial radius of $\sim 0.5R_{\odot}$   on some thermal timescale, $t_{\rm He, th.}$,  and engulf the NS initiating the second CE phase.  This timescale far exceeds the time it would take the NS to spiral-in and for the QN to be triggered.  Thus  $t_{\rm He, th.}$ is effectively the time between the occurrence of the first CE ejection and the time of the QN explosion.    Furthermore, the  propagation time of the ejected He CE to the H envelope is negligible because $v_{\rm H} << v_{\rm QN}$    where $v_{\rm H}$ is the  expansion speed of  the H envelope ejected during the first CE. It means that    when the second CE phase begins,  the previously ejected H-rich material will be located at a radius $R_{\rm H} \sim v_{\rm H} t_{\rm He, th.} \sim v_{\rm QN} t_{\rm CSM} $.

To model the third component, the CSM interaction in our model, we use the analytical bolometric light curve model of \citet{moriya_2013}. These models assume a constant CSM (i.e. wind) velocity $v_{\rm w}$  (here $v_{\rm w}=v_{\rm H}$) and a CSM density profile $\rho_{\rm CSM} = D r^{-s}$ where $D$ is a constant. The corresponding mass-loss rate is  $\dot{M}_{\rm CE, H} = \rho_{\rm  CSM} v_{\rm w} 4\pi r^2 = D v_{\rm w} 4\pi r^{2-s}$ with the $s=2$ case  corresponding to the steady mass-loss scenario  where $D= \dot{M}_{\rm CE, H}/4\pi v_{\rm w}$. The He-rich ejecta is defined by its kinetic energy $E_{\rm CE, He}\sim (1/2) M_{\rm He} v_{\rm QN}^2$. It has a double power-law profile for the density of homologously expanding ejecta ($\rho_{\rm ej}\propto r^{-n}$ outside and $\rho_{\rm ej}\propto r^{-\delta}$ inside). Another parameter in these models is the conversion efficiency from kinetic energy to radiation, $\epsilon$.

\begin{table*}[t!]
\begin{center}
\caption{Sample parameter fits to the  LCs  of double-humped SLSNe}
 \label{table:slsne-params}
\begin{tabular}{|c|c||c|c||c|c|c||c|}\hline
   \multicolumn{2}{|c||}{SLSNe}   & \multicolumn{2}{|c||}{QN} &  \multicolumn{3}{|c||}{BH Accretion}  & \multicolumn{1}{|c||}{Data shift$^{\dagger}$}  \\
  \hline
Name (restframe $\lambda$) & redshift & $M_{\rm CE}$ ($M_{\odot}$) & $R_{\rm CE, 0} (R_{\odot})$   &  $t_{\rm BH, delay}$ (days) & $L_{\rm 0}$ (erg/s) & $n_{\rm BH}$ &  days    \\\hline
PTF09cnd (4952\AA) & 0.258  & 2.8 & 800 &  12.0 & $4.2\times10^{44}$ &  0.4 & 3.0   \\\hline
SN1000$+$0216 (1535\AA) & 3.899 & 2.5 &3500 &   3.0 & $2.2\times 10^{45}$ & 0.1 & 4.0  \\\hline
PS1$-$10ahf (2890\AA) & 1.100 & 2.0 & 35 &    4.0& $1.5\times 10^{44}$ &  0.3 &  3.0 \\\hline
iPTF13ajg (3580\AA) & 0.740 & 2.3 & 180 &  7.5 & $8.0\times 10^{44}$ &  1.0 & 5.5  \\\hline
SNLS 06D4eu (2407\AA) & 1.588 & 2.3 & 60 &   5.2 &  $2.2\times 10^{45}$ &  1.2 & 2.0  \\\hline
PS1$-$10pm (2840\AA) & 1.206 & 1.8 & 45 &   12.0 & $1.2\times 10^{45}$  &   1.5 & 0.5   \\\hline
\hline
 iPTF13ehe (4639\AA) &  0.343 & 6.0  & 2000   &  22.0  &  $5.5\times 10^{44}$ & 1.1 &  3.0   \\\hline
 LSQ14bdq (3213\AA) &  0.345 & 2.8  & 900   &  15.5  &  $1.3\times 10^{44}$ & -0.4 &   3.0    \\\hline
\end{tabular}\\
$^{\dagger}$ Time of explosion in our model relative to $t=0$ in \cite{nicholl_2015}.
 \end{center}
\end{table*}

\section{Case studies}

\subsection{iPTF13ehe and LSQ14bdq}

We first fit the LCs of iPTF13ehe and LSQ14bdq  using the three-component model
described above.   Table 1 gives the best fit parameters for iPTF13ehe and LSQ14bdq\footnote{For LSQ14bdq, the numbers are slightly different from those given in Table 1 in \cite{ouyed_2015a} since we did not include its late-time emission.}. In the left panel in Figure \ref{fig:fits} our fit for iPTF13ehe in the r-band absolute magnitude
is compared to the observations from \cite{yan_2015}. The right panel shows the
fit for LSQ14bdq in g-band absolute magnitude with the observational data from \cite{nicholl_2015b}. The blue dashed curves shows the QN contribution (the first peak ). The black dotted
curve is the BH accretion and  the dot-dashed green curve shows the CE ejecta colliding with the CSM.

In the first light-curve component, best fits are obtained with  the QN going off when the He-rich CE  has
reached a radius of a few hundreds to a few thousands solar radii. 
 The second component is the BH-accretion phase (from merging of the QS with the CO core) occurring a few weeks following the QN proper.   For the third component, the CSM interaction
in our model,  for each fit, we use $n=12$, $\delta = 0$ and $\epsilon = 0.1$ while varying the other parameters  (see Table 1).  We are always in  a regime with $t<t_{\rm t}$ where   $t_{\rm t}$  is the time when the interacting region reaches the inner ejecta which is of the order of a few hundred days in our case (see eq. (8)
in \cite{moriya_2013}).   The best fit for the late-time lighcurve (i.e. the interaction with the H-rich material
from the first CE phase) is found for $t_{\rm CSM}= 225$ days and 190 days for iPTF13ehe and LSQ14bdq, respectively.
It is the time when the energized He-rich CE catches up with the H-rich material
from the first CE phase.  This means that the H-rich ejecta has expanded to 
a distance of $R_{\rm H}\simeq v_{\rm QN} t_{\rm CSM}\sim 10^{17}$ cm when the QN occurs. A distance 
reached in  a time of  $R_{\rm H}/v_{\rm H}\sim 100$ years  for $v_{\rm H}$ of a few hundred kilometers per second.
This timescale is effectively the time it would have taken the He-rich core to expand thermally and 
engulf the NS

We recall that when the  He core looses its hydrogen envelope  it finds itself out of equilibrium and will expand
  on thermal timescale $t_{\rm He, th.}$ to eventually engulf the NS to trigger the second CE.
  The  He-core will expand from an initial radius of $R_{\rm He} \sim 0.5R_{\odot}$
  to engulf the NS  (at distance of  $\sim 3R_{\odot}$) on a thermal timescale $t_{\rm He, th.}$.
 Using the virial theorem, we can estimate the thermal energy as $E_{\rm He, th.}\sim GM_{\rm He}^2/6R_{\rm He}$
  with  $G$ being the constant of gravity.  With a luminosity of $\sim 5\times 10^5 L_{\odot}$ (e.g.
  \cite{salaris_2005}) we get  $t_{\rm He, th.}\sim  10^3\ {\rm years}\times M_{\rm He, 8}^2/R_{\rm He, 0.5}$
  with $R_{\rm He, 0.5}$ and $M_{\rm He, 8}$ the He core's radius and mass in units of $0.5R_{\odot}$ and $8M_{\odot}$,
  respectively.   This would give a H envelope expansion radius of $\sim v_{\rm H} t_{\rm He, th.}\sim 10^{18}$ cm when it
  is hit by the He-rich ejecta,  an order of magnitude larger than the one inferred from the fit. 
  This suggests that the thermal timescale  is overestimated in the simple picture presented above.

\subsection{The QN peak in other double-humped SLSNe Ic}
\label{sec:otherslsne}

 \begin{table}[b!]
\begin{center}
\caption{Derived quantities}
  \label{table:slsne-derived}
\scalebox{0.8}{
\begin{tabular}{|c||c|c||c||c|}\hline
   \multicolumn{1}{|c||}{SLSNe}   & \multicolumn{2}{|c||}{Initial post-shock CE} &  \multicolumn{1}{|c||}{First hump}   & \multicolumn{1}{|c|}{$\eta_{\rm BH}M_{\rm CO}c^2$}  \\
  \hline
Name  &  $v_{\rm QN}$ ($10^4$ km/s)   & $T_{\rm CE, 0}$ ($10^6$ K) &  $T_{\rm eff.,  peak}$ ($10^3$ K) &  ($10^{50}$ ergs) \\\hline
PTF09cnd    & 2.17 & 1.06  & 17.0 & 4.7 \\\hline
SN1000$+$0216    & 2.30 & 0.35&  27.5 & 83 \\\hline
PS1$-$10ahf   &  2.57 & 11.1 & 24.3 & 1.2\\\hline
iPTF13ajg    & 2.40 & 3.24 &  19.3 & 0.32 \\\hline
SNLS 06D4eu    & 2.40 & 7.39& 28.3& 0.12 \\\hline
PS1$-$10pm   &  2.71 & 9.17& 21.7& 0.071 \\\hline
\hline
 iPTF13ehe      &  1.48 & 0.53 & 13.7 & 1.1\\\hline
 LSQ14bdq    &  2.17 & 0.97 & 17.3 & 19 \\\hline
\end{tabular}}\\
 \end{center}
\end{table}

\begin{figure*}[t!]
\centering
\includegraphics[scale=0.63]{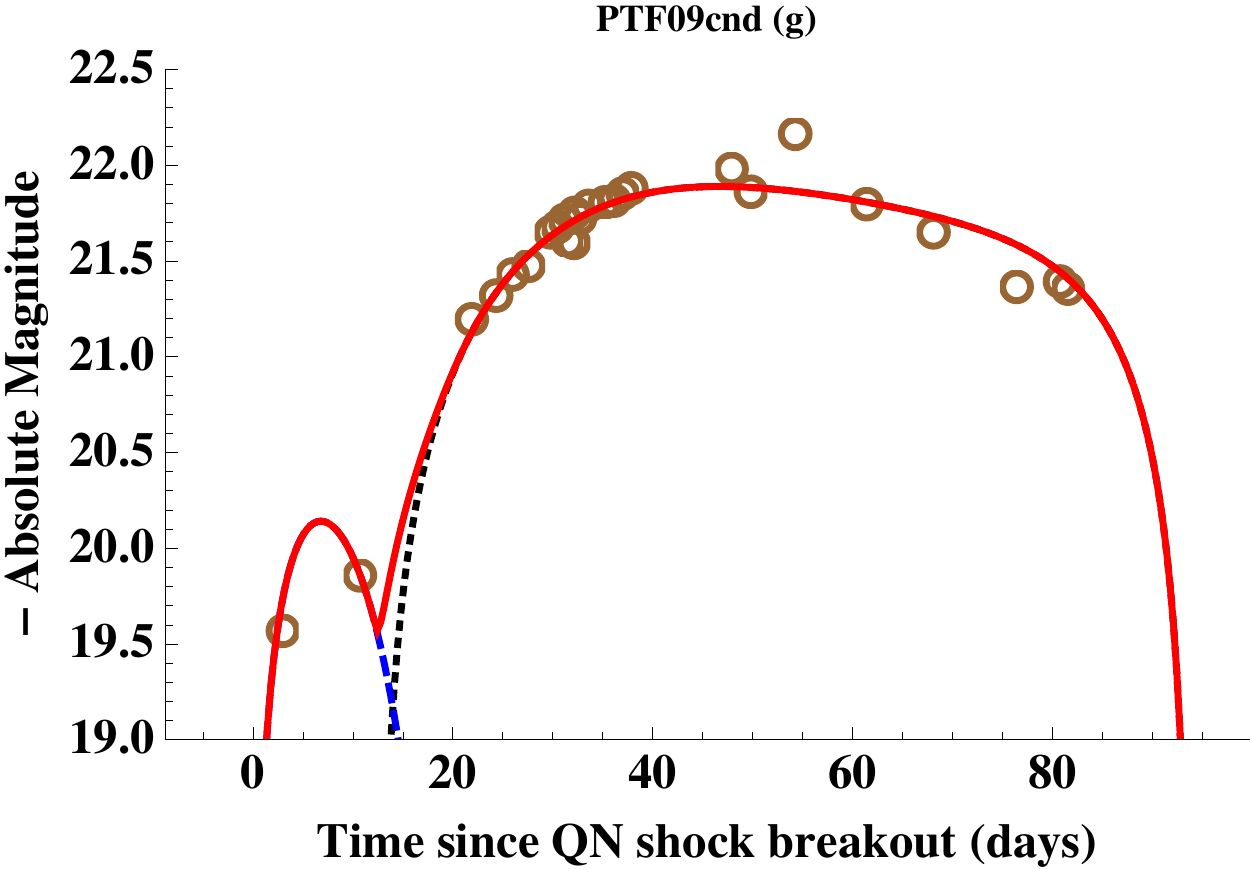}
\includegraphics[scale=0.63]{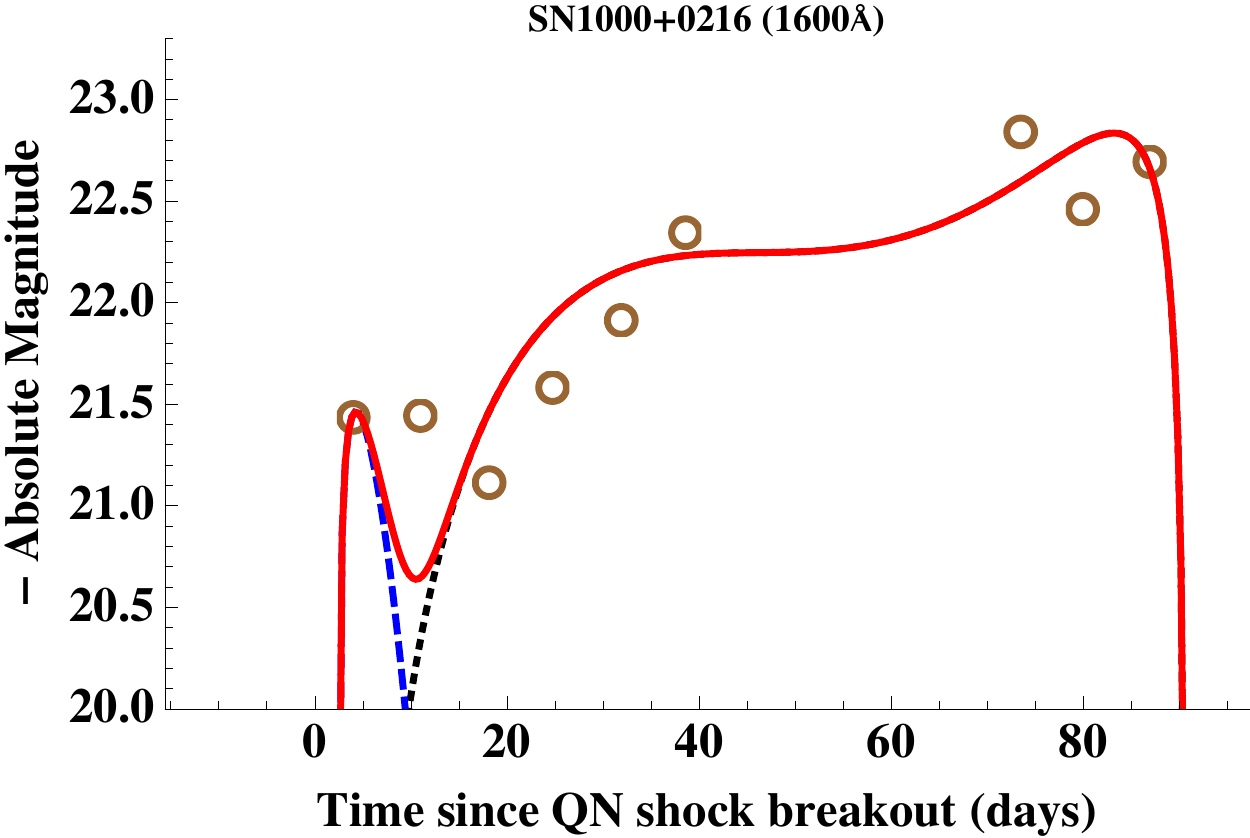}\\~\\
\includegraphics[scale=0.63]{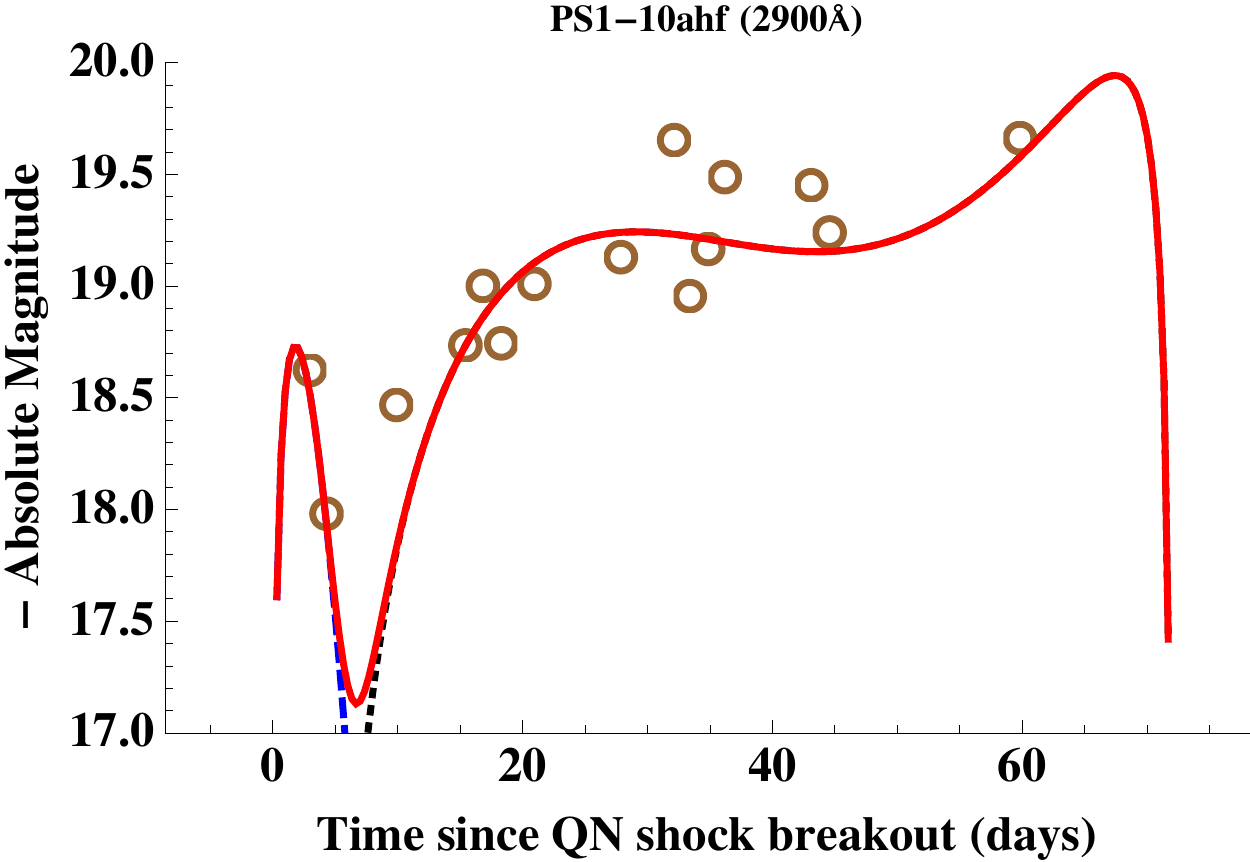}
\includegraphics[scale=0.63]{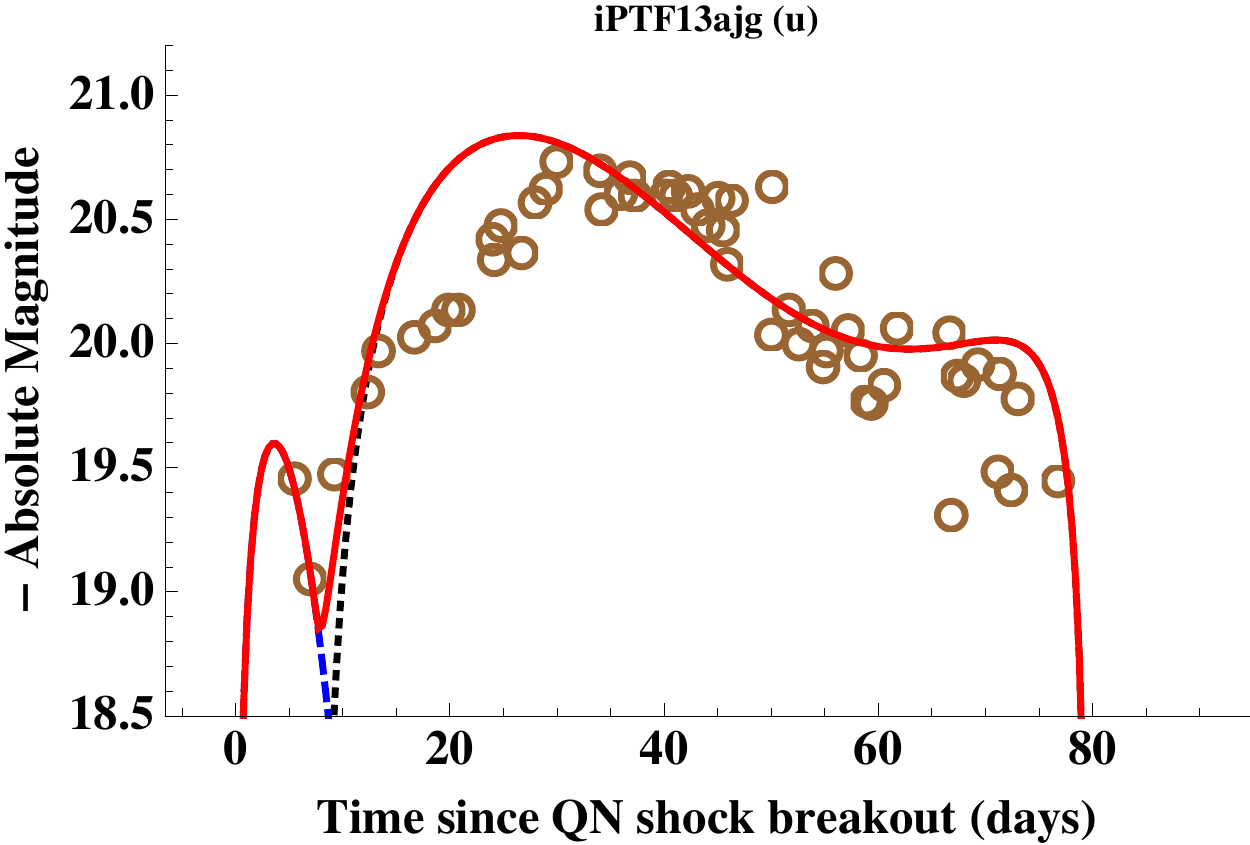}\\~\\
\includegraphics[scale=0.63]{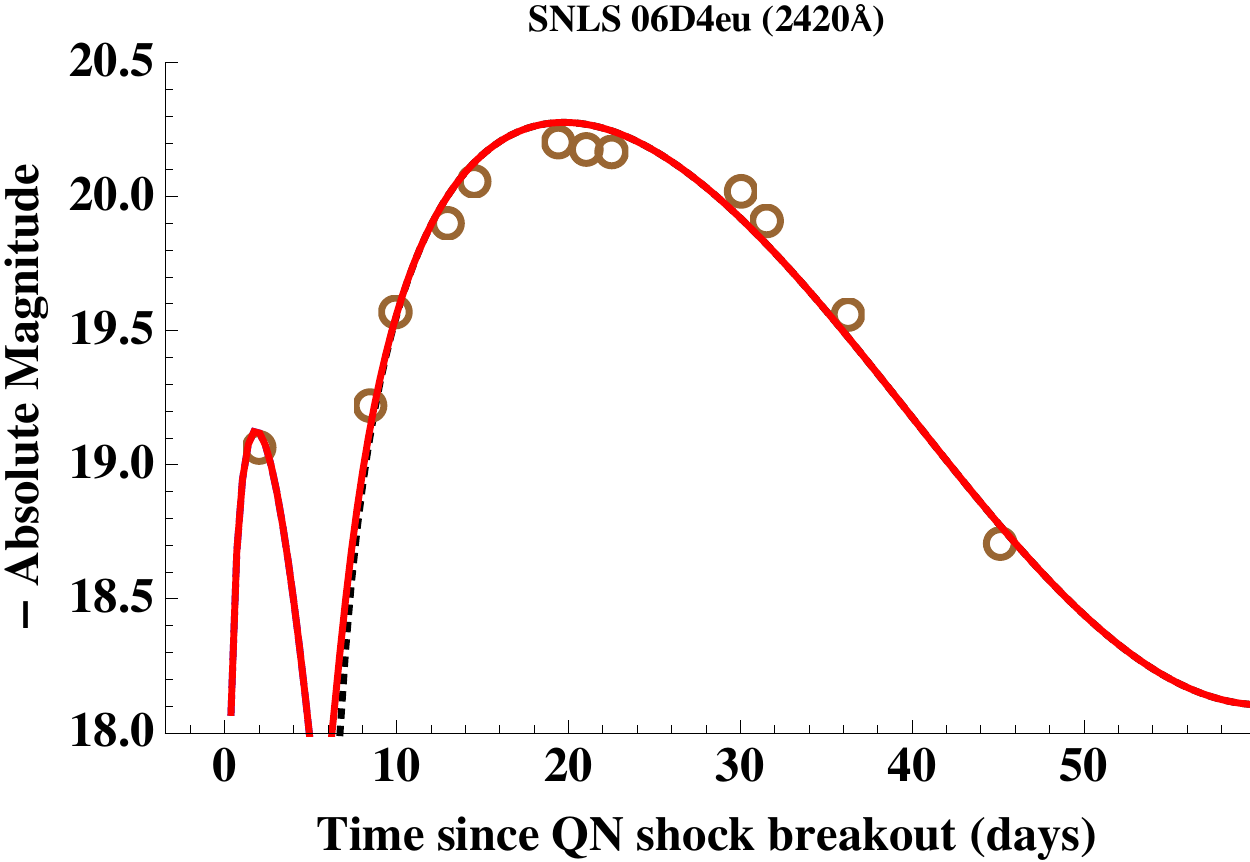}
\includegraphics[scale=0.63]{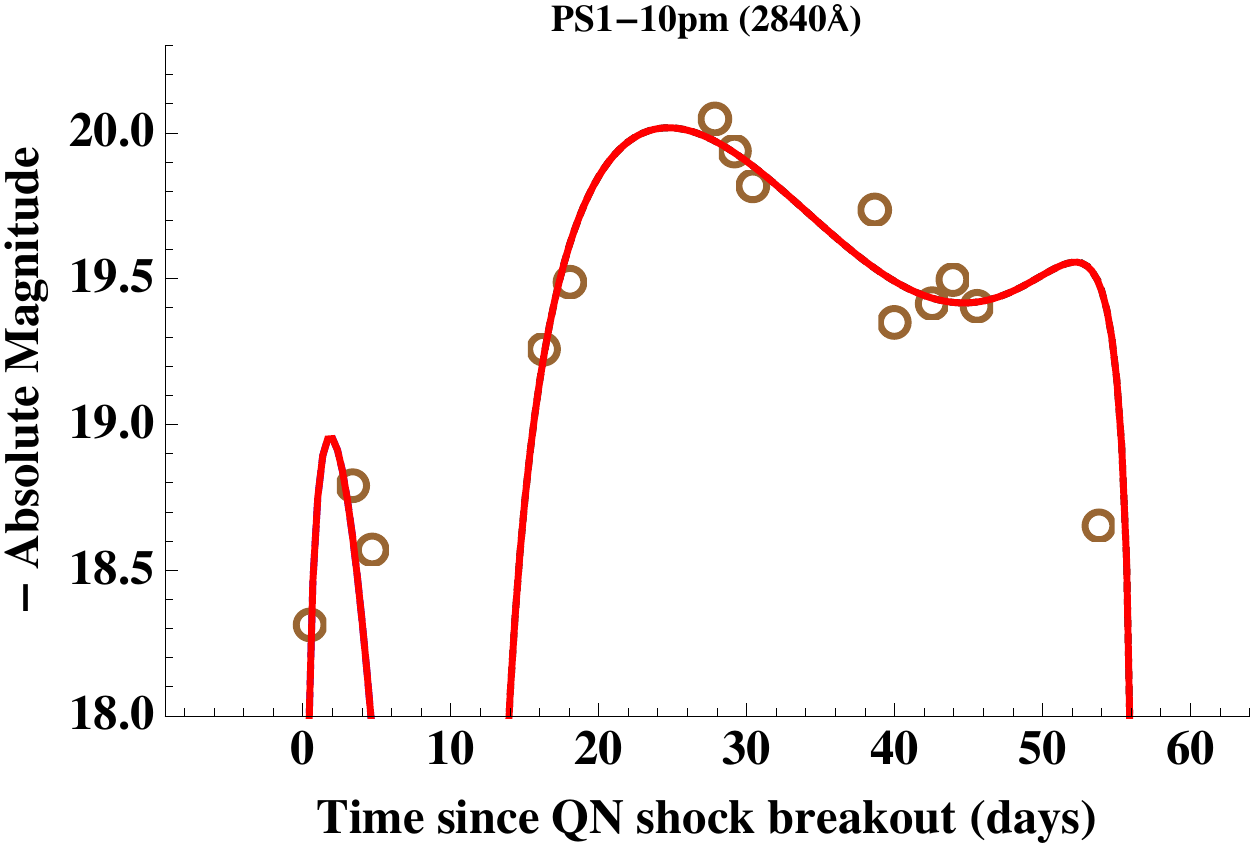}\\~\\
\includegraphics[scale=0.63]{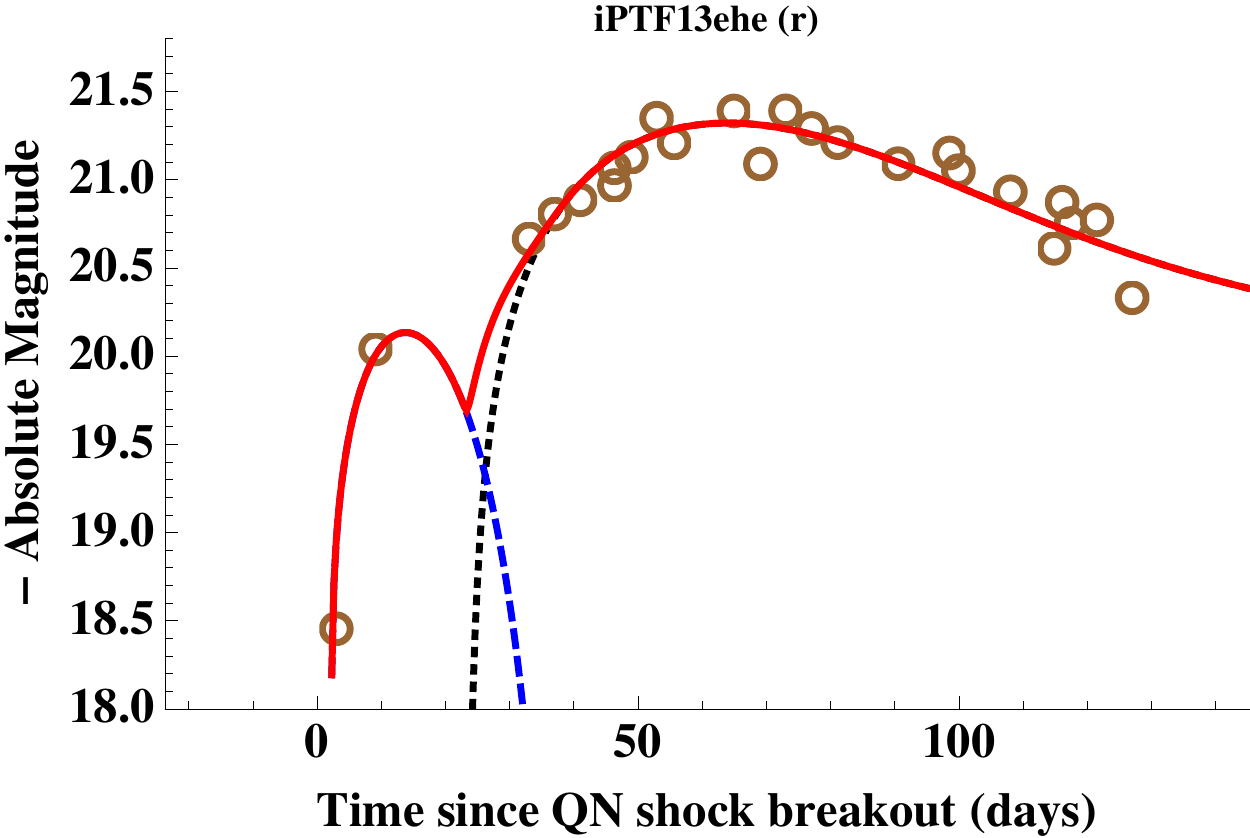}
\includegraphics[scale=0.63]{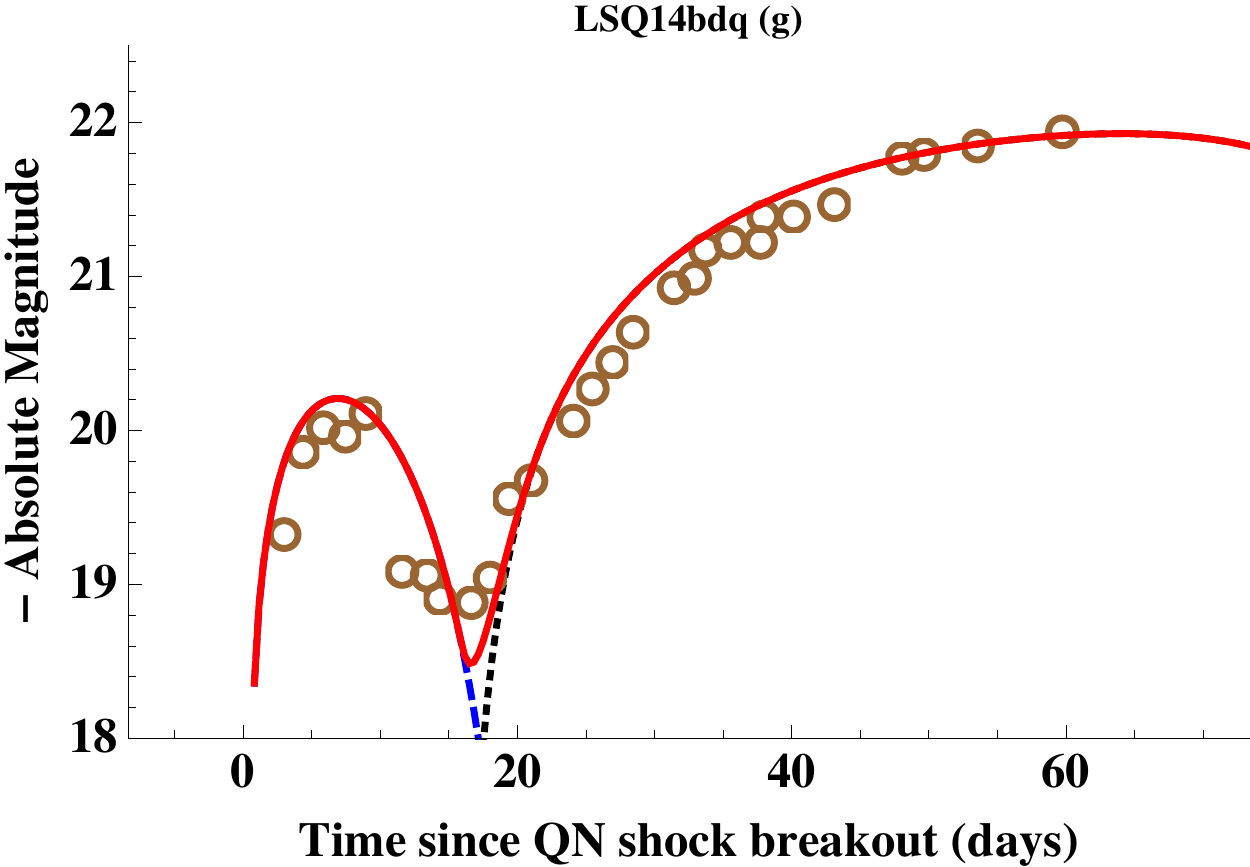}
  \caption{The QN model fit (solid red line) to six double-humped SLSNe Ic studied in Nicholl\&Smartt (2015); iPTF13ehe and LSQ14bdq are shown again for comparison  in the two bottom panels.  In all panels, the blue dashed curve  is the QN-powered (first) hump while the black dotted curve  is the BH-accretion powered (second) hump.}
\label{fig:slsne}
\end{figure*}

We now apply our model to six more SLSNe Ic showing hints of a double-humped
lightcurve as studied in \cite{nicholl_2015}.  Figure \ref{fig:slsne} shows our model's fits
using the parameters listed in Table  \ref{table:slsne-params};  iPTF13ehe and LSQ14bdq are included for comparison  in the two bottom rows.  For the first hump, we find  that the main
difference between these SLSNe in our model is due to the size, $R_{\rm CE, 0}$,   of the CE 
 when the QN goes off ($M_{\rm CE}$ shows much less scatter; i.e. a factor of a few compared to a factor of a hundred for $R_{\rm CE, 0}$). The fifth  column in Table \ref{table:slsne-params} shows the time delay between the BH accretion
phase and the onset of the QN explosion which varies from a few days to a few weeks while $n_{\rm BH}$ indicates
a range of injection power varying from rapidly decreasing to slowly increasing in time.

Table \ref{table:slsne-derived} shows the initial shock temperature  of the He-rich CE, $T_{\rm CE, 0}$, 
derived from $E_{\rm QN}$  and the corresponding ejecta velocity (see  \S \ref{sec:qn}). The fourth  column shows the effective temperature at the time of peak of the QN hump. Because the rest-frame wavelength varies, the peak
is observed earlier and with higher temperature for the higher redshift objects. The
temperature at peak varies significantly from $\sim 13,000$ K to $\sim 28,000$ K due to 
 these redshift effects  and  the variation in   $R_{\rm CE, 0}$ (cooler for large radii; comparing Table  \ref{table:slsne-params} and Table \ref{table:slsne-derived},  a large $R_{\rm CE, 0}$ results in a lower $T_{\rm CE, 0}$).
 Also shown, in the last column, is the   {\it minimum}  BH-accretion energy ($\eta_{\rm BH}M_{\rm CO}c^2$)
derived using Eq. (\ref{eq:dtBH}) with $\Delta t_{\rm BH}$ the observed (minimum) duration of the second hump in the data.  For BH accretion efficiency $\eta_{\rm BH}\sim 10^{-3}$ and $M_{\rm CO}\sim 2M_{\odot}$, Eq.(\ref{eq:dtBH}) gives a BH-accretion phase lasting a  few hundred days  which is more than adequate to provide the minimum required energy.

As noted in \cite{ouyed_2015a,ouyed_2015b}, when the first two peaks are distinguishable, the first peak's spectrum should be dominated by He lines while the second, BH-accretion powered, peak could additionally have spectral signatures of a CO-core composition deposited in the He-rich CE by the BH jet. The core composition when BH-accretion is triggered should be visible when the He envelope turns optically thin.

    The Thompson scattering mean-free-path of the H envelope 
    is $\sim 1/(n_{\rm H} \sigma_{\rm Th.})\sim 10^{19}$ cm where  $n_{\rm H}\sim N_{\rm H}/V_{\rm H}$
  with $N_{\rm H}= M_{\rm H}/(\mu_{\rm e} m_{\rm H})$ and $V_{\rm H} = (4\pi/3) R_{\rm H}^3$;  we assume 
    an electronic mean-molecular weight of $\mu_{\rm e}=2$ and a H envelope mass $M_{\rm H}\sim 20M_{\odot}$. 
    Thus it  will be transparent to photons from the QN and BH-accretion events.
  However, during the CSM interaction we expect  hard X-ray  emission from  the shock and 
an optical contribution from the fast-moving optically thick cooling  behind the shock which should lead to the emergence of a  H line in the late-time spectra (and lightcurves) of SLSNe.

Roughly 10-15\% of this type of massive binaries undergo the first CE phase when component B already has a CO
core. For the rest we still expect a QN hump and a possible CSM hump depending
on the thermal timescale of the stripped component B. In many cases we do not expect the
CSM interaction because the thermal timescale is longer for the cases without a CO core
(i.e. because of the lower luminosity). The H envelope will be located much further away and
much more dilute making the CSM interaction much fainter. The corresponding SLSNe-I should
show no Hydrogen and no He-burning signatures in their spectra.

\section{Conclusion}
\label{sec:conclusion}

 We demonstrate that a QN occurring in a massive binary which experienced 
two CE phases can explain  the double-peaked light curves of SLSNe-I.
If NSs in high-mass binaries can accrete sufficient mass to reach $M_{\rm NS, c.}$,
 a QN can result. In our model, the QN occurs when the CE has expanded to a large radius which yields the first peak in the SLSNe light-curve. The second peak is attributed to a
BH-accretion phase which occurs after the QS turns into a BH (as the QS merges with the CO core after
the ejection of the He-rich CE. The interaction of the ejected He-rich CE,
energized by both the QN explosion and the BH jet, with the H-rich material from the first CE phase
explains the bright late-tail emission and the corresponding broad $H_{\alpha}$ emission
observed in iPTF13ehe and LSQ14bdf.

  The possible use of  Type Ic  SLSNe as standardizable candles has been put forward in the literature
 (e.g. \cite{quimby_2013,inserra_2013,inserra_2014,scovacricchi_2015,papadopoulos_2015,wei_2015}).
Here, we have considered the double-humped subset and find that the first hump is best fit by considering a universal energy injection (i.e. the QN kinetic energy) into a variable CE envelope.  The 
connection between the first hump and the second  hump in our model through  $v_{\rm QN}, M_{\rm CE}$ and $R_{\rm CE,  0}$ is a natural outcome.  However, it is premature to infer that double-humped SLSNe Ic are standardizable  based on our model. This interesting topic will be explored further in the future.

 The success of our model at fitting light-curves of hydrogen-poor SLSNe (see {\it http://www.quarknova.ca/LCGallery.html}) hints that QNe do indeed occur in high-mass binaries as described in this work. We also
suggest that in these cases, QNe  assist with the CE ejection. We hope that this model can be tested
by inclusion of QN in simulations of CE evolution  which could also confirm that the
kinetic energy of the QN ejecta is a universal constant giving vital informations about the QN dynamics
and energetics.

\begin{acknowledgements}   
This work is funded by the Natural Sciences and Engineering Research Council of Canada. 
We thank the referee for helpful comments.
\end{acknowledgements}



\end{document}